\documentclass{article}

\usepackage{PRIMEarxiv}

\usepackage[utf8]{inputenc} 
\usepackage[T1]{fontenc}    
\usepackage{hyperref}       
\usepackage{url}            
\usepackage{booktabs}       
\usepackage{amsfonts}       
\usepackage{nicefrac}       
\usepackage{microtype}      
\usepackage{ulem}
\usepackage{etoolbox}
\usepackage[autostyle, english = american]{csquotes}
\usepackage{graphicx}
\usepackage{authblk}
\usepackage{amsmath, amsfonts}
\MakeOuterQuote{"}

\title{Neurophysiological correlates to the human brain complexity through $q$-statistical analysis of electroencephalogram}

\author[1,*]{Dimitri Marques Abramov}
\author[1]{Daniel de Freitas Quintanilha}
\author[2]{Henrique Santos Lima}
\author[1]{Roozemeria Pereira Costa}
\author[1]{Carla Kamil-Leite}
\author[1]{Vladimir V. Lazarev}
\author[2,3,4]{Constantino Tsallis}

\affil[1]{Instituto Nacional da Saude da Crianca, da Mulher e do Adolescente Fernandes Figueira, Fundacao Oswaldo Cruz, Avenida Rui Barbosa 716, Flamengo, Rio de Janeiro 22250-020, Brazil}
\affil[2]{Centro Brasileiro de Pesquisas Fisicas and National Institute of Science and Technology for Complex Systems, Rua Xavier Sigaud 150, Rio de Janeiro 22290-180, Brazil}
\affil[3]{Santa Fe Institute, 1399 Hyde Park Road, Santa Fe, New Mexico 87501, USA}
\affil[4]{Complexity Science Hub Vienna, Metternichgasse 8, 1030 Vienna, Austria}
\affil[*]{dimitri.abramov@fiocruz.br}

\begin{document}
\maketitle

\begin{abstract}
The prospects of assessing neural complexity (NC) by $q$-statistics of the systemic organization of different types and levels of brain activity were studied. In 70 adult subjects, NC was assessed via the parameter $q$ of $q$-statistics, applied to the ongoing and EEG and its spectral power of 20 scalp points (channels). The NC were estimated both globally for all channels (AllCh) and locally (for each single channel) in different Functional States (FSs). The values of $q$ was compared among FSs and single channels, as well they were correlated with the power of $\theta$ (4-8Hz),  $\beta_1$ (15-25Hz) and others EEG bands, in each FS. The value of $q$ across all FSs was higher for AllCh than for the single channels FSs. Consistently with previous studies, we found a negative correlation between NC and age. The FSs did not influence the $q$ of the EEG in AllCh, although locally the FS modulated $q$ in a consistent manner (e.g., reducing $q$ in posterior sites with eyes closed). The $q$ was correlated positively with the power of the $\theta$ and negatively with that of the $\beta_1$ band in general. These findings support the idea that, as a first approach, $q$-statistics can describe the human NC. The relationship between $q$ and $\theta$ power aligns with greater NC during FSs such as listening music and resting with eyes open, which is consistent with high-order representations rather than low-informative attentional tasks (OddBall).
\end{abstract}


\section*{Introduction}

Life is a phenomenon that emerges from complexity. Life begins when millions of different types of organic and inorganic molecules form an intricate network of short- and long-range interactions, from which self-sustaining endothermic dynamic states emerge, such as metabolism \cite{Tsallis1985, Yates1987}. Thus, complex systems such as living organisms manifest different levels of organization, ranging from the biomolecular networks, passing through epithelial or blood cells to diversified ecosystems where numerous species interact and evolve \cite{Maayan2017}. Despite universal properties, states at one level do not necessarily reflect the behavior of the system at a more global one. Neural networks form connective clusters around connective clusters yielding successive hierarchical levels of organization, where locally emerging neural codes of feature or action representations, which are integrated to manifest higher-order representations in larger scales, until a level of information organization in which subjective meanings in conscious experience ($qualias$, \cite{Kanai2012}) would be represented. The dynamics of these representations occur by coupling the activity of neurons distributed throughout the brain through phase locking between low and high frequencies, studied in EEG signals \cite{Lisman2013,Lundqvist2024,Karakas2020}.

Complexity relates to connectivity and informativeness in the system. It is a property of many natural and artificial systems, physically defined by long-range correlations between the components of these systems, which form an integrated and irreducible network while these components maintain their singularity \cite{Bassett2011}. Complex systems are therefore subject to chaotic fluctuations arising from the singularities of their components, which evolve through self-organization into order patterns that determine new states of the system as a whole, given its connectivity \cite{Kelso1995, Bassett2011}. The components of an ideal gas (molecular particles), although they manifest countless chaotic fluctuations all the time, do not establish long-range correlations, and therefore the probability of exhibiting macroscopic-order patterns becomes physically zero. Boltzmann and Gibbs developed statistical mechanics to model the probability of transformations of a system like the ideal gas, more probable as they reflect local interactions corresponding to low energies \cite{Boltzmann1872, Gibbs1902}, according to the function
\begin{equation}
   y = a/ \exp{(b\,x)} \;\;(b>0)
   \label{exponential}
\end{equation}
where the function (in this case, exponential) of the probabilities of events is determined by the parameters $a$ and $b$ (the normalization requirement determines naturally $a$ as a function of $b$ ). According to these probability functions, large-scale transformations (i.e., the emergence of collective states of order), which mobilize large amounts of energy, are extremely unlikely. However, this is not a statistical rule in nature, because macroscopic and stable manifestations of order involving large quantities of matter (on the order of $10^{23}$ molecules) and energy are quite common. Complexity explains these phenomena, as long-range interactions integrate the singular components of the system. In this context of integrality/singularity, it is not possible to analyze the components separately for a full understanding of the system composed of them. They are, in fact, irreducible. The irreducibility of complex systems requires non-additivity of the entropies (information) of their components in relation to the total entropy (information) of the system \cite{Tsallis2004, Tsallis2023}. Entropic non-additivity is a fundamental characteristic of complex systems, where the whole differs from the sum of its parts. In 1988, Tsallis developed a mathematical generalization of Boltzmann and Gibbs’ Statistical Mechanics (BGSM), which models the dynamics and thermal behavior of systems, whether complex or not, more adequately than BGSM \cite{Tsallis1988}: non-extensive statistical mechanics (NESM), or $q$-statistics. In its $q$-generalised Gaussian or exponential functions, the index $q$ determines the kurtosis of the probability function. This index $q$ is, in some way, a measure of entropy and, by corollary, of information. Thus, $q$ quantifies complexity, as demonstrated in the literature. A more general expression emerges, namely, the following $q$-exponential function:
\begin{equation}
 y = a/[1+(q-1)b\,x]^{(1/(q-1))} \,,
 \label{qexponential}
\end{equation} 
which recovers Eq. (\ref{exponential}) for $q=1$ and where the normalization requirement determines $a$ as a function of $(b,q)$. The functions of NESM have consistently proven suitable for modeling the dynamics of complex natural, artificial and social systems and characterising their complexity through $q$ \cite{Tsallis1988, Tsallis2004, Tsallis2023}. However, some biological and social systems have shown singular statistical behavior, where the highest probabilities of events/transitions do not appear at the lowest energies \cite{Mitsokapas2021, Picoli2009, Tsallis2003, Tsallis2020, Upadhyaya2001, Wild2023}. In this case, exponential distributions can have a maximum and can be modeled by a $q$-exponential function with a prefactor such as $a\,x^c$. We demonstrated the application of $q$-statistics to describe the statistical behavior of the human brain through the electroencephalogram (EEG) signal, using the following function \cite{NOSSO2023}:
\begin{equation}
  y=a\,x^c/ [1 + (q-1) b\, x^h]^{(1/(q-1))} 
  \label{eq3}
\end{equation}
which recovers Eq. (\ref{qexponential}) for $(c,h)=(0,1)$. 
This describes a probability distribution; the normalization requirement determines the parameter $a$ to be given by:
\begin{equation}
   a^{-1}= \left[ b \cdot (q - 1) \right]^{-\frac{c+1}{h}} \cdot \frac{\Gamma\left(\frac{c+1}{h}\right) \cdot \Gamma\left(\frac{1}{q-1} - \frac{c+1}{h}\right)}{h \cdot \Gamma\left(\frac{1}{q-1}\right)} \,.
   \label{eq4}
\end{equation}
 In this case, $c$ determines the slope of the left tail (low energies) and could be explained by the degeneracy of the states of the physical system (the number of co-occurring states with the same energy, \cite{Reichl2005}), which, in statistical terms, is related to degrees of freedom. In our previous works \cite{NOSSO2023,NOSSO2024}, the complex behavior of the EEG was studied through the regularity of the signal, i.e., how much the signal presents itself as a stationary time series or alternatively a chaotic time series. This regularity can be described, for example, through autocorrelation or it can be measured through the frequency distribution of the time intervals between specific events in the EEG, such as amplitudes that exceed a pre-defined threshold \cite{NOSSO2023}. A highly regular signal (as a sinusoidal steady pattern) would correspond to a narrow frequency distribution, where almost all intervals would have the same duration, a signal with very low complexity. If the signal is a random time series, the distribution converges to the Boltzmann-Gibbs statistics. 
 
 Here, we have studied the distribution of frequencies of occurrence of time intervals between amplitudes that exceeded a threshold of -1.0 standard deviation. In addition to observing that $q$-statistics (and not BGSM) appropriately models the distributions of inter-event intervals in the EEG of fifteen typical adults \cite{NOSSO2023}, we presented the first evidence that $q$-statistics is sensitive and specific enough to differentiate the EEG of different mental profiles (in this case, children with ADHD and typical ones) through the couple of parameters $(q,c)$ \cite{NOSSO2024}. These profiles correspond to different types of EEGs at some level of their organization, thus denoting different neural complexities for different kinds of mental processing.

The EEG is the macroscopic phenomenon of the electrical dynamics resulting from the collective activity of neural populations, recorded through electrodes distributed across the scalp, forming capture channels, with each channel registering the local dynamics of the circuits immediately beneath it \cite{Niedermeyer2011}. The signal has a non-linear macroscopic aspect with unpredictable behavior at time $t$+1, but with well-defined order patterns, such as the Posterior Dominant Rhythm (PDR) or $\beta$ activity registered in the frontal channels \cite{Niedermeyer2011}. The EEG is the result of the vectors of the electrical activity of billions of cells, widely connected over short and long distances forming subsystems within subsystems according to connection density, across various levels of the organization \cite{Stam2012}. Therefore, it reflects the complexity of the brain as a whole, as well as these subsystems, when evaluating single channels.

The EEG is also a time series which can be mathematically modeled through numerous superimposed stationary waves, whose frequencies have amplitudes (power) and transient phases. These waves form a frequency spectrum, with powers and phases described, for instance, through Fourier Transform (FT) functions \cite{Niedermeyer2011}. Different frequency components can be related in certain ways to certain cognitive processes and mental events. For example, the characteristics of oscillations in the $\theta$ band (between 4 and 8 Hz) are associated with memory, learning \cite{Herweg2024}, voluntary attention and mental selectivity \cite{Lazarev2006}, and high cognitive performance \cite{Tan2024, Lisman2013, Karakas2020}. Therefore, it is possible that the EEG $\theta$ band, as a correlate of integrative information processing, can be correlated to NC.

The modulation of features of the EEG frequencies during performance of cognitive and behavioral tasks demonstrates certain patterns of emergent global states of the brain, related to respective mental and behavioral processes. These processes are dynamic and transient, and they are reflected in both temporal dynamics and spectral composition of the EEG as, for example, the manifestation of the PDR while eyes are closed \cite{Niedermeyer2011}. Different functional states may be reflected in different characteristics of the NC in the EEG, both globally (all channels) and locally (single channels), in a differentiated manner. Thus, we could explain the functional significance of NC and assign its different patterns to respective types of mental processes. 

In this study, we aimed to evaluate the sensitivity and informative value of the $q$-statistics in relation to the functional states of the brain. We explored the characteristics of NC in the less-regular part of EEG of typical adults across various states determined by different tasks and procedures in relation to the spectral composition of respective EEG signals, removing the Dominant Posterior Rhythm (PDR). It is a pattern of stereotypal activity reflecting thalamo-cortical synchronization between 8-12Hz, related to lowering information in visual system while the eyes are closed, some kind of "stand-by" cortical state\cite{Niedermeyer2011}. The PDR exhibits high amplitude super-regularity that biases the empirical distributions with a second peak of probability \cite{NOSSO2024}. When eyes are closed and PDR shows higher amplitude, the fitting is usually distorted. Preventing bias in the NC analysis of the closed eyes EEG, we had to suppress a narrow band of all empirical probability distributions, related to PDR.

In the less-regular part of EEG (without PDR), we expected to discriminate different FSs in global NC (AllCh) and local NC (single channels). We studied the following FSs: resting states of relaxed wakefullness with eyes open and closed, during intermittent photic stimulation of various frequencies, mental arithmetic and an Oddball test, which is a monotonous operational routine usually employed in the study of P300, a brain potential related to selective and sustained attention \cite{Duncan2009}. The last FS was accessed through listening to a preferred, emotionally significant piece of music previously selected by the subject, when we expect to evoke a multidimensional affective experience, compatible with the recruitment of multiple brain systems \cite{Vuust2022}. We also have accessed the relation between the age of the subjects and respective $q$ values, since studies points to negative effect of aging and diseases on complexity \cite{Hatterjee2017, Goldberger2002, Hong2007, Ishii2017, Lopez2023, Manor2012}. In addition, we verified differences in NC related to biological sex at the global level (AllCh).

\section*{Methodology}
\subsection*{Participants}\vspace{0.5cm}
Seventy typical adults (36 female, all cisgender) aged between 21 and 87 years were included, evenly distributed across both sexes. Participants were excluded if they: (1) had a current history of mental disorders or self-reported dysfunctional mental distress, (2) had a diagnosis of neurological diseases, (3) had a current history of uncompensated or untreated clinical illnesses (e.g., diabetes, hypertension), (4) had consumed alcohol or benzodiazepines in the last 48 hours or other psychotropic drugs in the last 30 days, (5) used psychotropic medications for any purpose, or (6) were pregnant. All participants provided their written informed consent to participate in the study after receiving adequate explanations and signing the corresponding consent form. This research was approved by the Research Ethics Committee of the Fernandes Figueira Institute (CEPIFF) under opinion number 5,874,944 and is registered under CAAE 66854323.4.0000.5269.

\subsection*{Experimental Procedures}\vspace{0.5cm}
Participants underwent a brief medical history assessment to evaluate their state of physical and mental health, medication use, lifestyle habits, and physiological rhythms (ovulatory, sleep, etc), which lasted approximately 15 minutes. Following this, participants completed three scales: the Adult ADHD Self-Report Scale (ARSR) with six questions for ADHD symptom screening \cite{Adler, Ustun2017}, the Adult Autism Questionnaire – 10 questions, to assess autism symptoms \cite{Auyeung2008}, and the Self-Perceived Stress Scale (PSS), quantifying subjective stress perception\cite{Reis2010}. The information from these scales and tests was not used in this study. The intelligence status (IQ) of the subjects was accessed by applying the 13 subtests of the Wechsler Adult Intelligence Scale, third ed. (WAIS-III) \cite{Wechsler}. 

Subsequently, EEG was recorded when subjects remained comfortably seated, keeping their body relaxed, in a climate-controlled experimental room (22 to 24 ° C) and acoustically attenuated. The EEG electrodes were placed manually on the scalp, plus one bipolar channel for the electrocardiogram (ECG) signal acquisition. Each participant underwent seven tasks/procedures corresponding to the seven different FS in EEG recording: (1) Resting state of relaxed wakefulness with eyes open free to visually explore under light for five minutes; (2) Resting state of relaxed wakefulness with eyes closed for 5 minutes, with the lights off; (3) Intermittent photic stimulation (IPS) of low frequencies by a white stroboscopic light, color 6500 K, 30 cd/m²/W, 20 µs pulse duration at sequencies of 4Hz, 7Hz and 11Hz, successively (120 s each) with eyes closed, and (4) IPS of high frequencies (15Hz and 25Hz), successively, also for 120 s each. Afterward, active tasks began (all with open eyes): (5) Mathematical (arithmetic) Operations, for five minutes, when we ask the participant to mentally solve multiplications  in the format 3 digits times 1 digit (e.g. 253 x 6)  as quickly as possible, pressing a button upon completing each calculation. (6) Attention Test (OddBall Paradigm), when the participant was presented with 450 short tones of low (500 Hz) and  high (1000 Hz) frequency ordered pseudo-randomly, with frequent (85$\%$) and rare (15$\%$) occurrence, played in PC speakers at an intensity of 50 dB at the participant's head position (40cm distance each sound box), and was asked to promptly identify the rare sound pressing a button. Each tone lasted 200ms, with a 1000ms interstimulus interval. Finally, (7) Preferred Music, when the participants listened to their favorite song, previously selected, for at least 5 minutes using headphones. The participant were instructed to choose songs that emotionally moved them. 

\begin{figure}[ht]
\centering
\includegraphics[width=\linewidth]{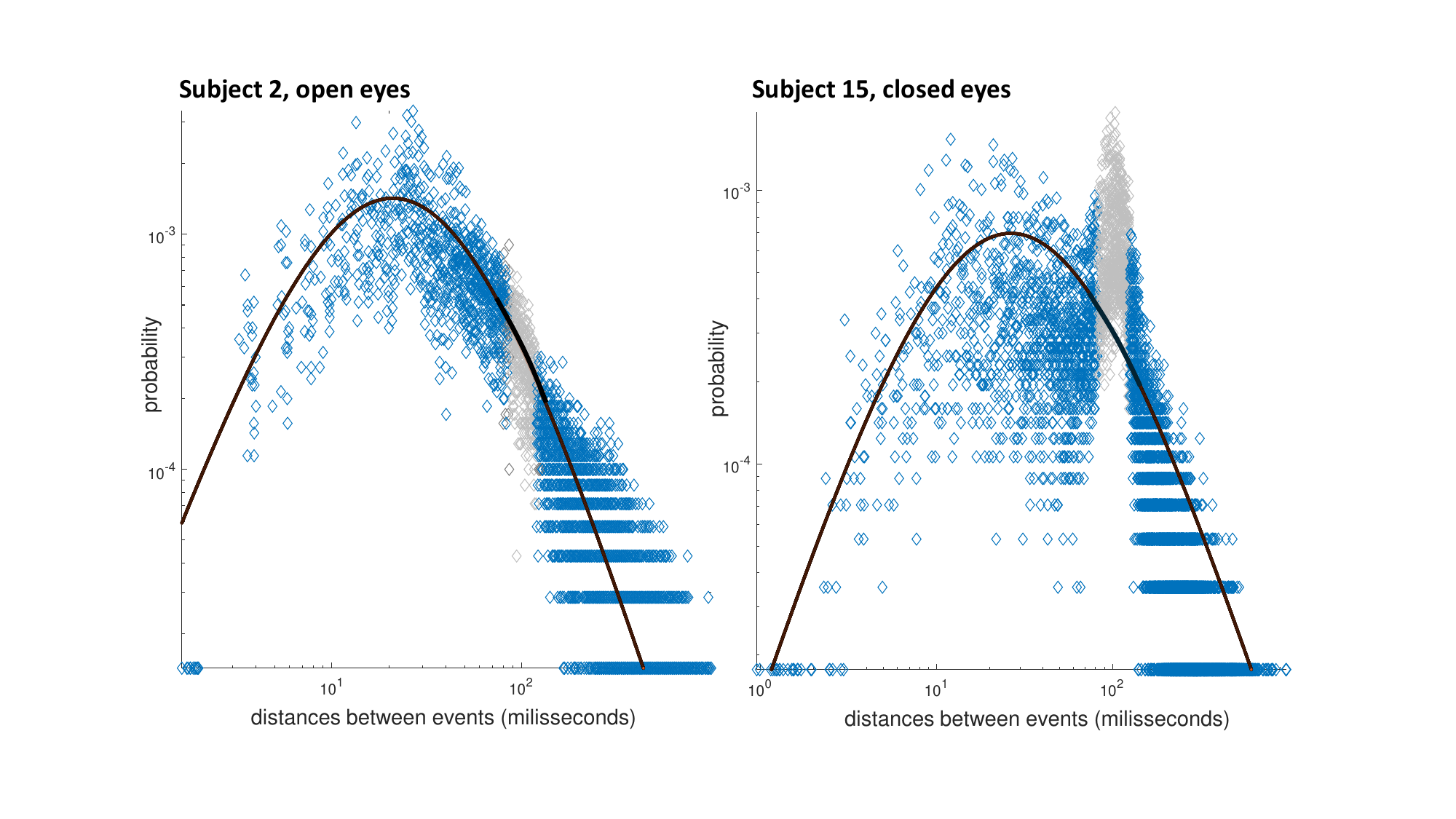}
\caption{Two examples of the $q$-exponential function (equation 3) fitted to empirical probability distributions of occurrence of event intervals collected from the EEG signal (all 20 channels). Each event is defined when the signal amplitude crosses down the threshold of -1.0 standard deviation of the negative part of the signal. The gray dots (event intervals between 80 and 120 ms) were removed from the fitting process because they correspond to Posterior Dominant Rhythm (8-12Hz) of the signal, which, in many distributions, forms a peak in probability (a highly regular pattern) that distorts the fitting (right side, resting with eyes closed). Parameters for the fitting in the distribution of Subject 2, in resting (eyes open): $b = 0.1463$, $c = 2.2473$, $h = 1.0792$, $q = 1.2237$; parameters for the fitting of Subject 15, in resting (eyes closed): $b = 0.1527$, $c = 2.0081$, $h = 1.0000$, $q = 1.2503$.}
\label{fig1}
\end{figure}

\subsection*{EEG Recording and Filtering}\vspace{0.5cm}
The EEG signal was recorded using a Nihon-Kohden NK 1200 device, through 20 acquisition channels positioned according to the International 10-20 System (including Oz), with reference to the linked mastoids (A1+A2). Data acquisition was performed with a sampling rate of 1000 Hz. Two electrodes of a bipolar channel were placed 5cm below the midpoint of each clavicle (positive on the left) for ECG signal acquisition.

The signal was filtered using a convolution function as seen in \cite{NOSSO2024}, which attenuated frequencies below 0.5Hz (high-pass) and above 100Hz (low-pass). Using an FFT filter, electrical network artifacts at 60Hz were suppressed. An automated algorithm was used to suppress EEG segments with amplitudes greater than 3 standard deviations from the mean of the signal, corresponding to muscle and electrostatic artifacts. When present, electrical artifacts from cardiac activity were also automatically suppressed using the ECG signal as a time reference. After all, the EEG signal was manually inspected to exclude remaining sections contaminated by muscle artifacts. Due to the potential distortion of the EEG signal by various artifacts, no algorithm was used to attenuate vertical electrooculogram artifacts (blinks), which have positive polarity.

\subsection*{Data Processing and Curve Fitting}\vspace{0.5cm}
Regularities were assessed through the frequency distribution of event intervals. These events were defined when the negative amplitude of the signal exceeded the threshold of $-1.0$ standard deviation of the negative part of the signal, truncated at amplitudes of $100 mV$, suppressing any remaining artifacts, as normal brain waves in awake adults
do not exceed the range of 100mV (-50 to +50mV) \cite{Niedermeyer2011}. Finally, the time segments excluded (replaced by $NaN$ in the signal vector) were suppressed.
Event intervals were distributed into 500 classes, ranging from 0 to 1000ms in steps of 2ms ($0,2,4,\dots ,998,1000$). Data from the classes between 80ms and 120ms corresponding to the PDR (8-12Hz) were suppressed.

The interval distributions between events were fitted to the distributions according to Eqs. \eqref{eq3},\eqref{eq4}  using a least squares method. For the parameters $(b, c, h, q)$, 
we used the lower bounds [0, 0, 1, 1.01] and upper bounds [0.5, 4, 2, 2], with starting points [0.02, 2.3, 1.1, 1.25] and a precision of the order of $10^{-10}.$ We determined the four parameters for each of the 20 channels (single channels) as well as for the 20 distributions consolidated into a single "cloud" (AllCh, as shown in Figure \ref{fig1}).
\begin{figure}[ht]
\centering
\includegraphics[width=\linewidth]{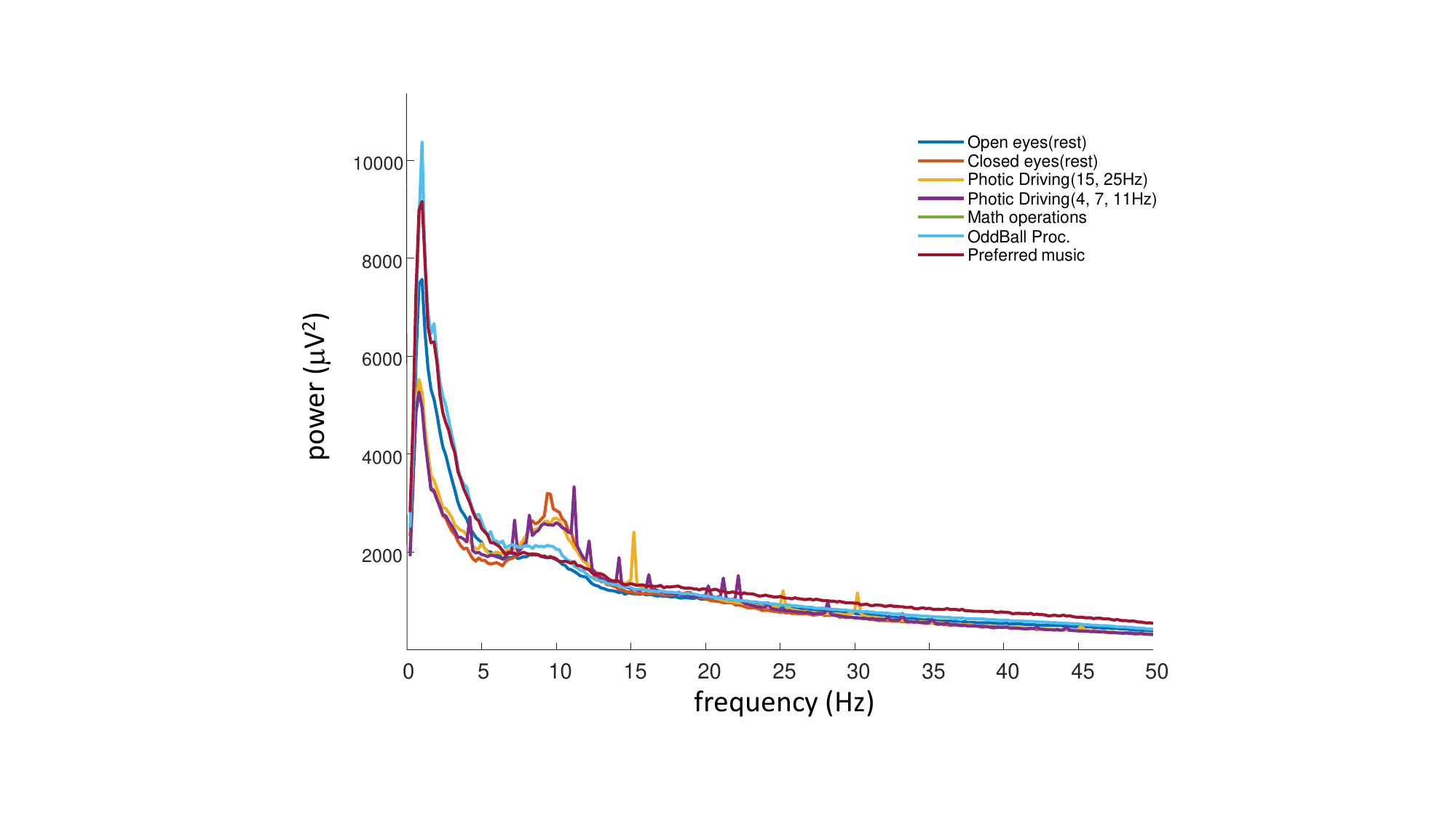}
\caption{Power spectra of the frequency signals (0 to 50Hz), averaged across all subjects for each functional state (FS) (legend), obtained from successive 5s windows of each EEG. From these spectra, we calculated the average power in the $\delta$ (0-4Hz), $\theta$ (4-8Hz), $\alpha$(8-12Hz), $\beta_1$ (15-25Hz), $\beta_2$ (25-35Hz), and $\gamma$ (35-50Hz) bands. Note that Intermitent Photostimulation caused photic driving, with secondary power peaks due to  harmonics and sub-harmonics (shown in orange and purple in the figure).}
\label{fig2}
\end{figure}

We calculated the RMSE (root mean square error) of the residuals from the least squares fitting, excluding fittings with a probability of error in convergence greater than $1\%$. For all channels, the result of each fitting was also inspected, and we excluded cases where there was no visible convergence at the left or right tails. The images of each distribution (by subject and state) for all channels, and each channel with its respective fittings, are available in Supplementary Information S1, RMSEs in Supplementary Table S2, and excluded cases by visual inspection in Supplementary List S3. 

We performed the EEG spectral analysis using the DFT (Discrete Fourier Transform). The signal for each subject and channel was divided into 5-second epochs, where the power of frequencies between 0 and 50Hz (a total of 250 frequencies) was computed. The power spectra from the epochs were averaged, resulting in an average spectrum for each channel from each subject. The average power for each band was computed from the average power between 1 and 4Hz ($\delta$), 4.19 and 8Hz ($\theta$), 8.19 and 12Hz ($\alpha$), 15 and 25Hz ($\beta_1$), 25.19 and 35Hz ($\beta_2$), and 35.19 and 50Hz ($\gamma$). The power spectra as a function of frequency for each FS studied (averaged across subjects) are shown in Figure \ref{fig2}.

\section*{Results}

\subsection*{Sample}\vspace{0.5cm}

Out of 70 participants, one did not complete the research procedures, and three were excluded due to extremely poor EEG signal quality caused by numerous artifacts. This left a total of 35 women (age: 47.03 $\pm$17.71, meaning $\pm$ standard deviation) and 31 men (age: 44.56 $\pm$ 16.61), with no significant statistical difference in age between genders ($p = 0.550$, t-test for independent samples).

Based on RMSE, no signals were excluded. Regarding all channels (AllCh), 35 EEG signals of specific FS were excluded due to not showing visual convergence of the q-exponential function (as per the methodology). Whenever relevant, the number of cases used in the analyses is stated in the tables, figures, or text.

During signal filtering, on average, $11.96\%$ of the signal from each FS of each subject was automatically or manually suppressed due to artifacts in blocks lasting at least 1 second (Supplementary Table S4, sheet 1). In signal processing, remaining artifacts (any amplitude below -100mV) accounted for only $0.0001\%$ of the EEG signal (Supplementary Table S4). Therefore, the interference from these suppressions in the spectral analysis and the construction of the event interval distributions was negligible.

The IQ of the subjects (assessed by WAIS) was 112.3 $\pm$9.90 (women) and 117.52 $\pm$ 12.38 (men), with no statistically significant difference ($p = 0.065$, t-test for independent samples). As expected, no correlation was found between age and IQ (women: $(r,p(r))=(0.14,0.406)$; men: $(r,p(r))=(0.08,0.627)$.

Participants were noticeably anxious during the mathematical calculations, as reflected in their speech and the increased occurrence of unwanted movements and muscular artifacts (qualitative observation). All participants maintained engagement and focus on the OddBall task, although they commonly considered it boring.

The IPS in any frequency caused Photic Driving FSs with primary and secondary power peaks, related to each IPS frequency and their harmonics (Figure \ref{fig2}). 

\subsection*{Complexity}\vspace{0.5cm}

We initially compared the average value of the $q$ parameter for the 20 single channels with the $q$ value obtained from all channels (AllCh). In all FS, the $q$(AllCh) was significantly higher than the average $q$ value of the single channels (Table 1).
\begin{table}[h!]
\centering
\begin{tabular}{|l|c|c|c|c|c|c|c|c|}
\hline
\textbf{EEG state} & \multicolumn{3}{c|}{\textbf{all channels}} & \multicolumn{3}{c|}{\textbf{mean of single channels}} &  &  \\ \hline
 & \textbf{N} & \textbf{mean} & \textbf{std.dev} & \textbf{N} & \textbf{mean} & \textbf{std.dev} & \textbf{t-stat} & \textbf{p-value} \\ \hline
resting(OE) & 66 & 1,196 & 0,053 & 66 & 1,176 & 0,047 & 2,75 & 0,008 \\ \hline
resting(CE) & 66 & 1,205 & 0,056 & 66 & 1,184 & 0,050 & 5,35 & 0,000 \\ \hline
ph.driving(low) & 66 & 1,186 & 0,060 & 66 & 1,168 & 0,046 & 3,89 & 0,000 \\ \hline
ph.driving(high) & 66 & 1,194 & 0,064 & 66 & 1,176 & 0,050 & 3,43 & 0,001 \\ \hline
mat. operation & 66 & 1,205 & 0,046 & 66 & 1,182 & 0,047 & 5,18 & 0,000 \\ \hline
oddBall-P300 & 66 & 1,196 & 0,049 & 66 & 1,175 & 0,043 & 3,52 & 0,001 \\ \hline
preferred music & 66 & 1,196 & 0,040 & 66 & 1,179 & 0,047 & 2,73 & 0,008 \\ \hline
\end{tabular}
\caption{Parameter $q$: Averaged channels vs All channels (t-test for paired samples)}
\label{table:1}
\end{table}

The value of $q$ (AllCh) did not depend on biological sex, with means for women and men of 1.19$\pm$0.05 and 1.20$\pm$0.06, respectively, in resting (eyes open) ($t = -1.13$, $p = 0.265$, t-test for independent samples); 1.20$\pm$0.05 and 1.21$\pm$0.06 in resting (eyes closed) ($t = -0.98$, $p = 0.329$); 1.18$\pm$0.06 and 1.20$\pm$0.05 in Photic driving (low frequency) ($t = -1.28$, $p = 0.204$); 1.19$\pm$0.07 and 1.20$\pm$0.05 in Photic driving (high frequency) ($t = -0.90$, $p = 0.372$); 1.20$\pm$0.04 and 1.21$\pm$0.05 in math operations ($t = -1.03$, $p = 0.305$); 1.19$\pm$0.05 and 1.21$\pm$0.04 in OddBall ($t = -1.52$, $p = 0.133$); and 1.19$\pm$0.04 and 1.20$\pm$0.04 in preferred music ($t = -0.96$, $p = 0.344$).

\begin{figure}[ht]
\centering
\includegraphics[width=\linewidth]{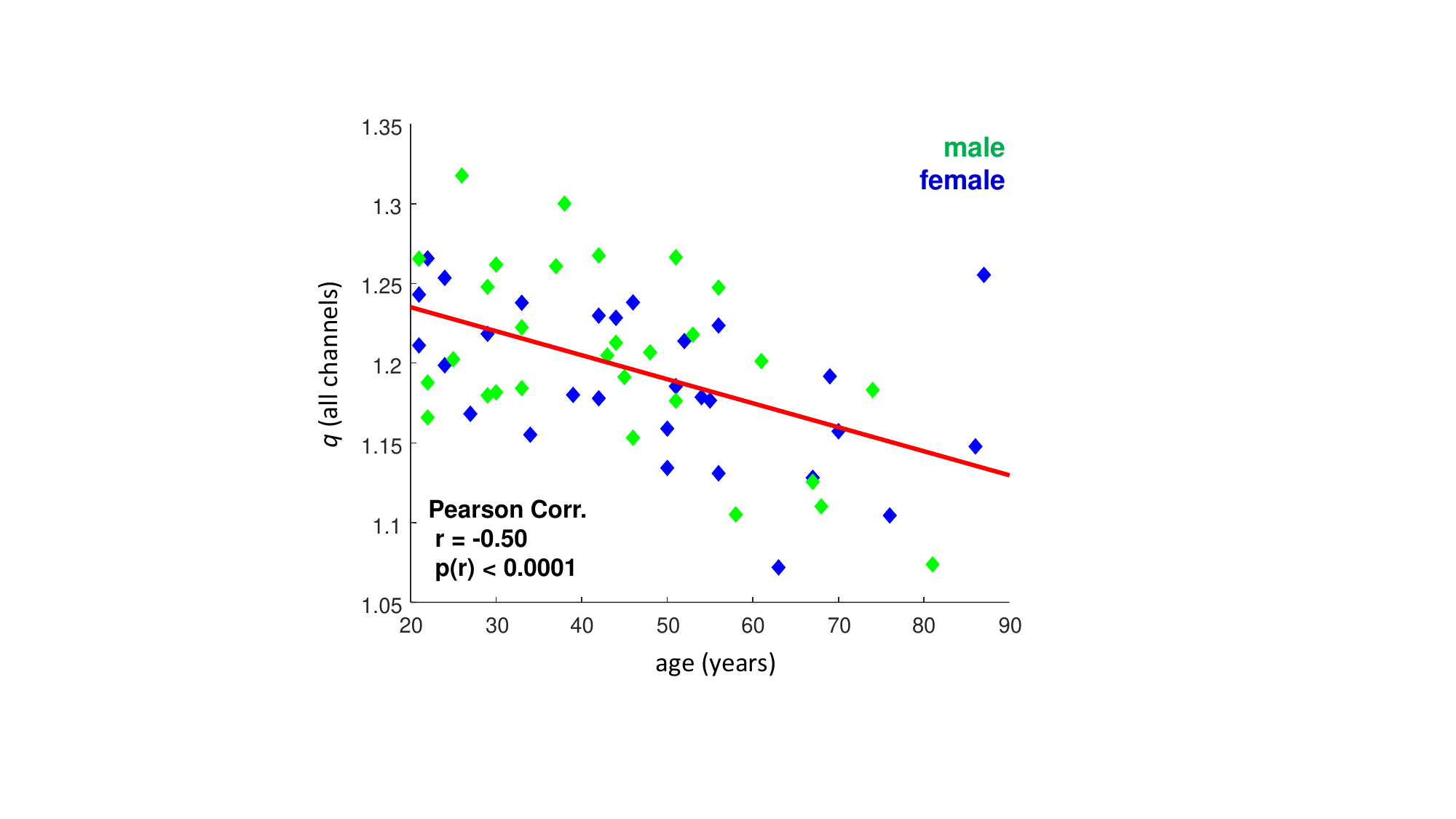}
\caption{The ages of the subjects negatively correlate to respective $q$ values for all channels. Data from EEG in the resting (open eyes) mental state.  All of the sex dichotomies are merely illustrative. }
\label{fig3}
\end{figure}

\begin{table}[h!]
\centering
\begin{tabular}{|l|c|c|c|}
\hline
\textbf{State} & \textbf{n} & \textbf{r} & \textbf{p(r)} \\ \hline
resting(OE) & 58 & -0,50 & 0,000 \\ \hline
resting(CE) & 62 & -0,22 & 0,083 \\ \hline
ph.driving(low) & 62 & -0,22 & 0,081 \\ \hline
ph.driving(high) & 62 & -0,34 & 0,006 \\ \hline
math. operations & 59 & -0,28 & 0,031 \\ \hline
oddBall-P300 & 60 & -0,34 & 0,008 \\ \hline
preferred music & 57 & -0,33 & 0,012 \\ \hline
\end{tabular}
\caption{Age x NC (Pearson Correlation), All channels}
\label{table:pearson}
\end{table}

The values of $q$ were negatively correlated with age (Figure \ref{fig3}, in the resting state (eyes open), and Table \ref{table:pearson} in all seven FS) ($r$ = -0.50, $p < 0.0001$, $n$ = 58). The negative correlation was not significant only for the eyes closed and photic driving (low frequencies) states ($r$ = -0.22 in both cases), with the highest correlation coefficient observed for resting (open eyes) ($r$ = -0.50) and the lowest for the mathematical operations state ($r$ = -0.28).

We correlated the age of the subjects with the value of $q$ in each state and each single channel, producing 140 Pearson correlations. We observed that only three were not negative but with $r$ coefficients $< 0.13$ (not significant). All other correlations were negative, ranging from $r=-0.004$ to $r=-0.63$, with 88 of the 140 correlations ($62\%$) being statistically significant ($p < 0.05$, data in Supplementary Table S5). 

\begin{figure}[ht]
\centering
\includegraphics[width=\linewidth]{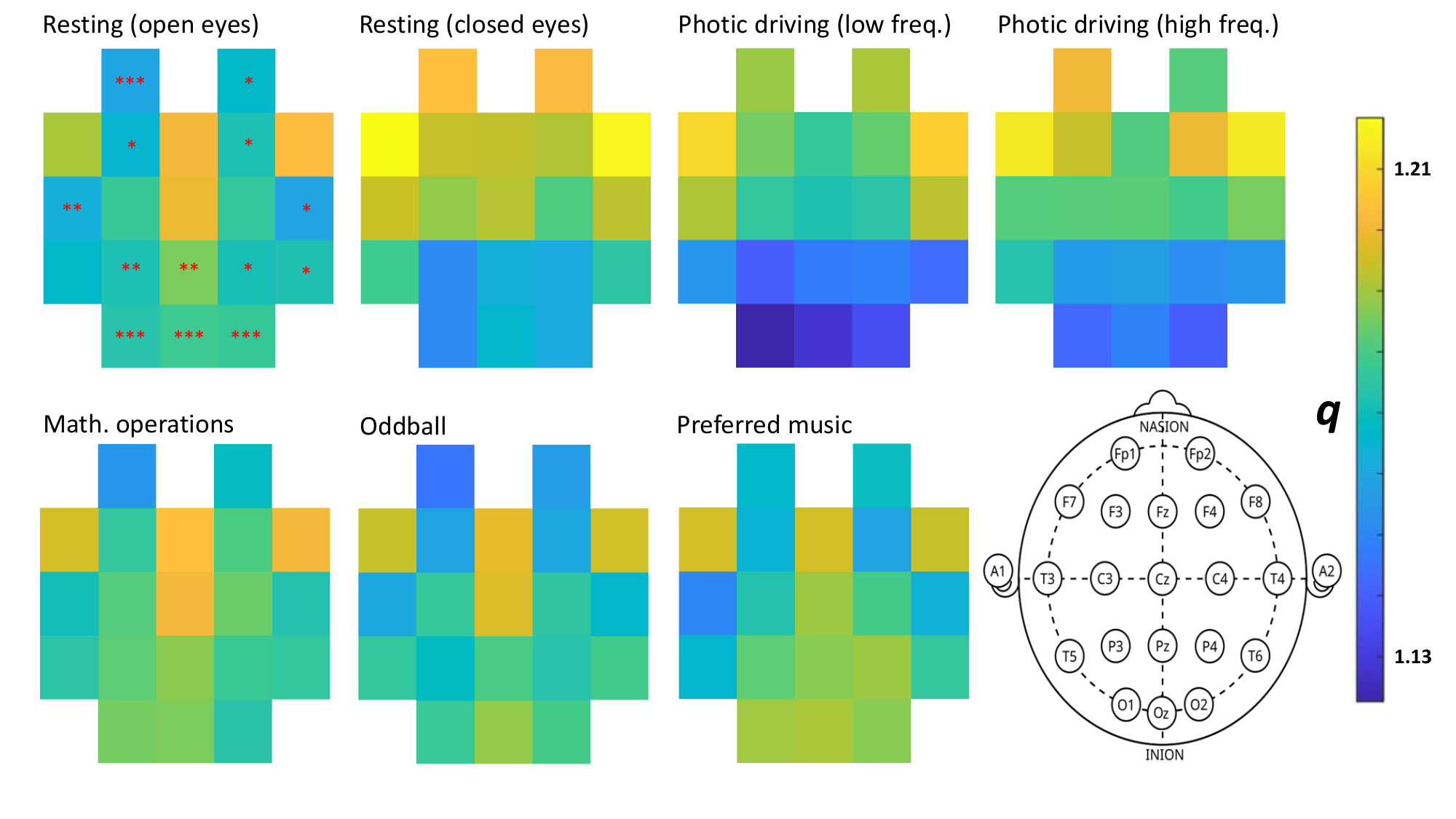}
\caption{Effect of the functional state (FS) on the $q$ parameter in single channels. The level of $q$ is encoded by the color map on the vertical bar to the right, and the squares in the colored plots represent each of the 20 EEG channels on the scalp, with reference to A1+A2 (placement scheme inserted at the bottom right of the figure). In the plot for the FS Resting (open eyes), the probability of the absence of an effect of the FS on the value of $q$ is encoded (One Way ANOVA, *: $p < 0.05$, **: $p < 0.01$, ***: $p < 0.001$).}
\label{fig4}
\end{figure}

\begin{figure}[ht]
\centering
\includegraphics[width=15cm]{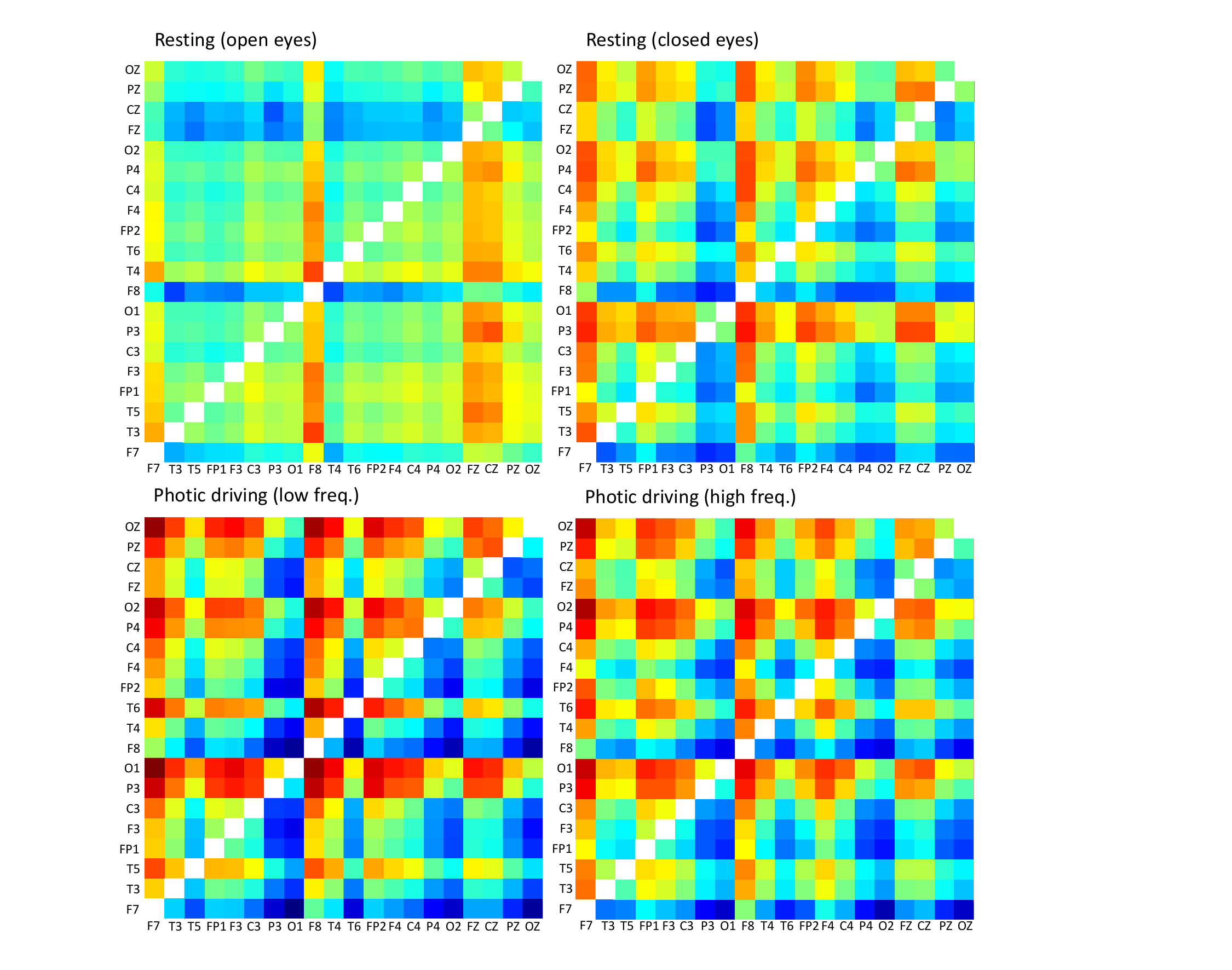}
\caption{The t-values (t-test) for the paired comparisons of the $q$ parameter values between channels (column $\times$ row) in Functional States (FS) at resting (eyes open and closed, with or without photic driving). The color scale (bottom right) encodes the t-values. Positive t-values indicate that the channel (column) $>$ channel (row), and the opposite for negative t-values. The colorbar indicates the range of t-values corresponding to p $\ge$0.05. (see Figure \ref{fig6})}
\label{fig5}
\end{figure}

\begin{figure}[ht]
\centering
\includegraphics[width=15cm]{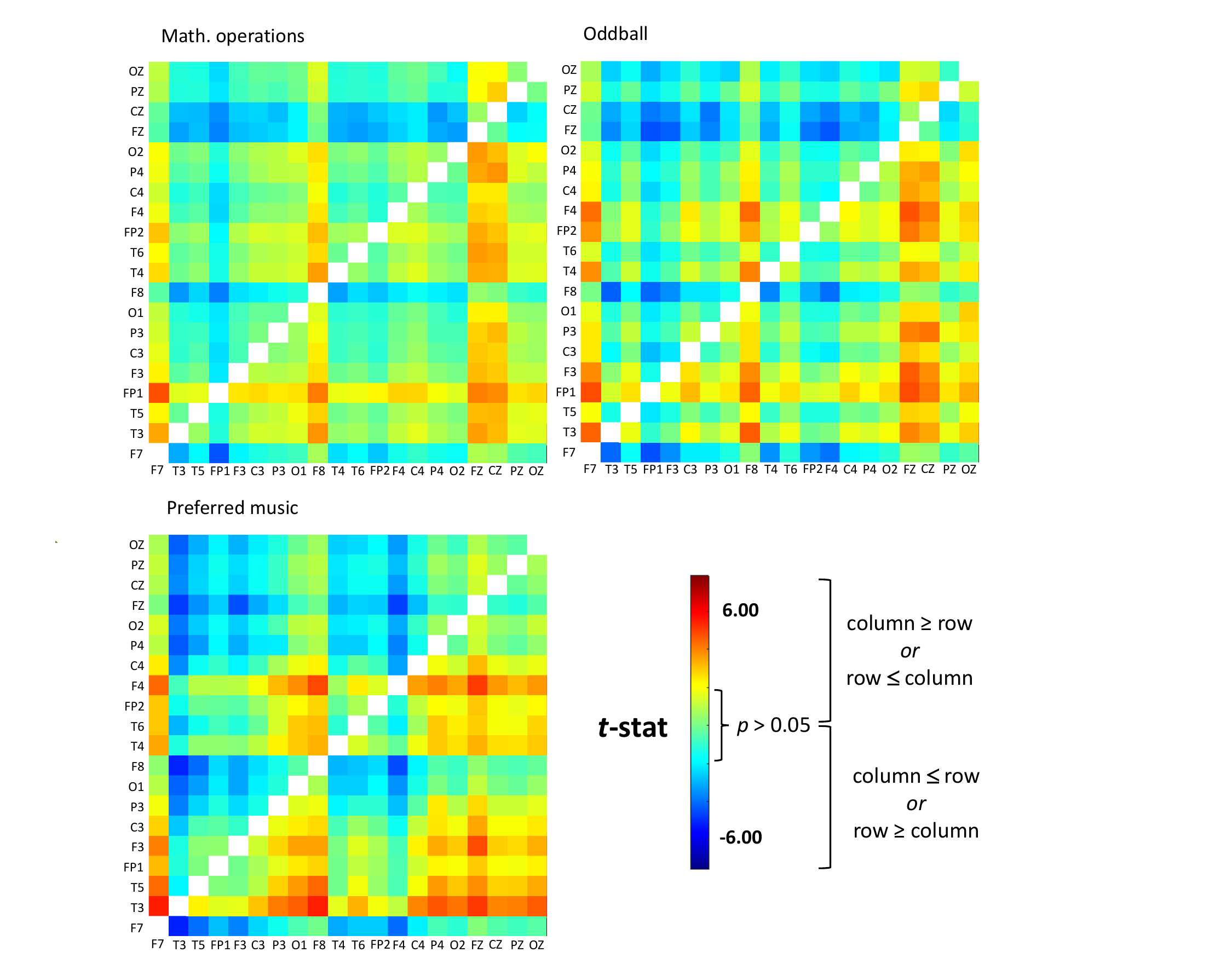}
\caption{The t-values (t-test) for the paired comparisons of the $q$ parameter values between channels (column $\times$ row) in Functional States (FS) Mathematical operations, oddball and preferred music (see Figure \ref{fig5} caption)}
\label{fig6}
\end{figure}

Through analysis of variance, no effect of the FS on the values of $q$ (all channels) was found ($F = 0.90$, $DF=6$, $p = 0.491$, Table 1). However, a noticeable effect was observed in single channels across the different FS (Figure \ref{fig4}). In Figure \ref{fig4}, we observe the effect of the FS on the value of $q$ in each channel independently, which was significant in the channels FP1, O1, OZ, and O2 ($p < 0.001$, one-way ANOVA), T3, P3, PZ ($p < 0.01$), and FP2, F3, F4, T4, and T6 ( $p < 0.05$, Figure \ref{fig4}, panel corresponding to the resting (open eyes) functional state, at the top left). Clearer variations were in the occipital parietal region (bottom center), where $q$ is less pronounced with eyes closed and more pronounced in the preferred music state. During eyes closed, $q$ in FP1 and FP2 is higher than in the eyes open states. In all states, we observed higher values of $q$ in the left (F7) and right (F8) anterior temporal channels, while in the median frontal (FZ) and central (CZ) channels, the differences in $q$ were not significant. Middle temporal  (T3 and T4) channels showed higher $q$ in all eye-closed states. 

We compared the value of the $q$ parameter between pairs of channels in each FS (Figures \ref{fig5} and \ref{fig6}, showing the t-values from the paired t-test). In resting (eyes open), the EEG at the fronto-central median and anterior temporal electrodes showed a higher value of $q$ (Figure \ref{fig5}). This is a pattern that somehow persisted across all states with eyes open.

In the resting state (eyes closed), there was a significant relative reduction in the complexity of the EEG in the occipital parietal (P3,P4, O1, O2) and posterior temporal(T5, T6) areas as well as in the median frontocentral region (F3,F4,C3,C4) (Figure \ref{fig5}).

The effect of photic driving appears to be similar for both low and high frequencies (Figure \ref{fig5}). Photic driving further reduced $q$ at the posterior electrodes compared to the others, maintaining reduced frontocentral complexity relative to the resting (eyes open) state.

The values of $q$ increased in the frontotemporal and frontocentral electrodes in cognitively demanding states (Figure \ref{fig6}, in mathematical operations and Oddball).

During the preferred music FS (Figure \ref{fig6}), we observed a relative increase in the value of $q$ at the posterior channels compared to the central and anterior ones. There appears to be a higher increase on the right side, highlighted by the $t$-statistics showing that $T4 > T3, T6 > T5$, and $P4 > P3$.

\begin{table}[h!]
\centering
\begin{tabular}{|l|c|c|c|c|}
\hline
\textbf{State} & \textbf{n} & \textbf{r} & \textbf{p(r)} & \textbf{Sig.} \\ \hline
delta (1-4Hz) & 58 & 0,30 & 0,0239 & \# \\ \hline
$\theta$ (4-8Hz) & 58 & 0,48 & 0,0001 & \#\#\# \\ \hline
alfa (8-12Hz) & 58 & 0,38 & 0,0036 & \#\# \\ \hline
beta1 (15-25Hz) & 58 & -0,25 & 0,0571 & \# \\ \hline
beta2 (25-35Hz) & 58 & -0,17 & 0,1944 & \\ \hline
gama (35-50Hz) & 58 & -0,14 & 0,2811 & \\ \hline
delta/beta1 & 58 & 0,46 & 0,0002 & \#\#\# \\ \hline
delta/beta2 & 58 & 0,38 & 0,0030 & \#\# \\ \hline
delta/gama & 58 & 0,31 & 0,0177 & \# \\ \hline
teta/beta1 & 58 & 0,69 & 0,0000 & \#\#\#\# \\ \hline
teta/beta2 & 58 & 0,55 & 0,0000 & \#\#\#\# \\ \hline
teta/gama & 58 & 0,44 & 0,0005 & \#\#\# \\ \hline
alfa/beta1 & 58 & 0,70 & 0,0000 & \#\#\#\# \\ \hline
alfa/beta2 & 58 & 0,51 & 0,0001 & \#\#\#\# \\ \hline
alfa/beta3 & 58 & 0,40 & 0,0020 & \#\# \\ \hline
\end{tabular}
\caption{Pearson Corr. Between $q$ (all channels), open eyes, and respective EEG bands}
\label{table3}
\end{table}

\subsection*{Spectral Composition and Complexity}\vspace{0.5cm}

The power spectra as a function of frequency exhibit a typical profile, with the highest absolute power in the $\delta$ band, an increase in power in the $\alpha$ band (mainly during FS with eyes closed) and a gradual reduction of power with increasing frequency. See Figure \ref{fig2}, where we visually noted that the average spectra of each FS are slightly different. We especially observed narrow and discrete power peaks at the frequencies of intermittent photo-stimulation showing Photic Driving FSs, as well as their harmonics (e.g., a power peak at 4Hz that is the stimulation frequency, and harmonics at 8, 12, 16, and 24Hz, purple wave), and sub-harmonics (e.g., at the stimulation frequency of 15Hz, sub-harmonic at 5Hz, and harmonic at 30Hz, orange wave). This myriad of resonances in the form of harmonics interferes with the observation of specific photic driving effects at low and high frequencies. It also introduces a bias in the correlation between the spectral composition of the EEG signal and the $q$ parameter (we did not analyzed correlations between power spectra and photic driving FSs).

The Pearson's correlations between absolute ($\delta$, $\theta$, $\alpha$, $\beta_1, \beta_2$, and $\gamma$) and relative ($\theta/\beta_1$ or $\alpha/\beta_1$) band powers with $q$ parameter were studied for both AllCh and each single EEG channel. The average power in AllCh showed a significant positive correlation between the low-frequency bands and the $q$ values, as well as non-significant negative correlations (with a highlight on $r=-0.21$ with $\beta_1$) with the high frequencies (Table \ref{table3}).

\begin{figure}[ht]
\centering
\includegraphics[width=\linewidth]{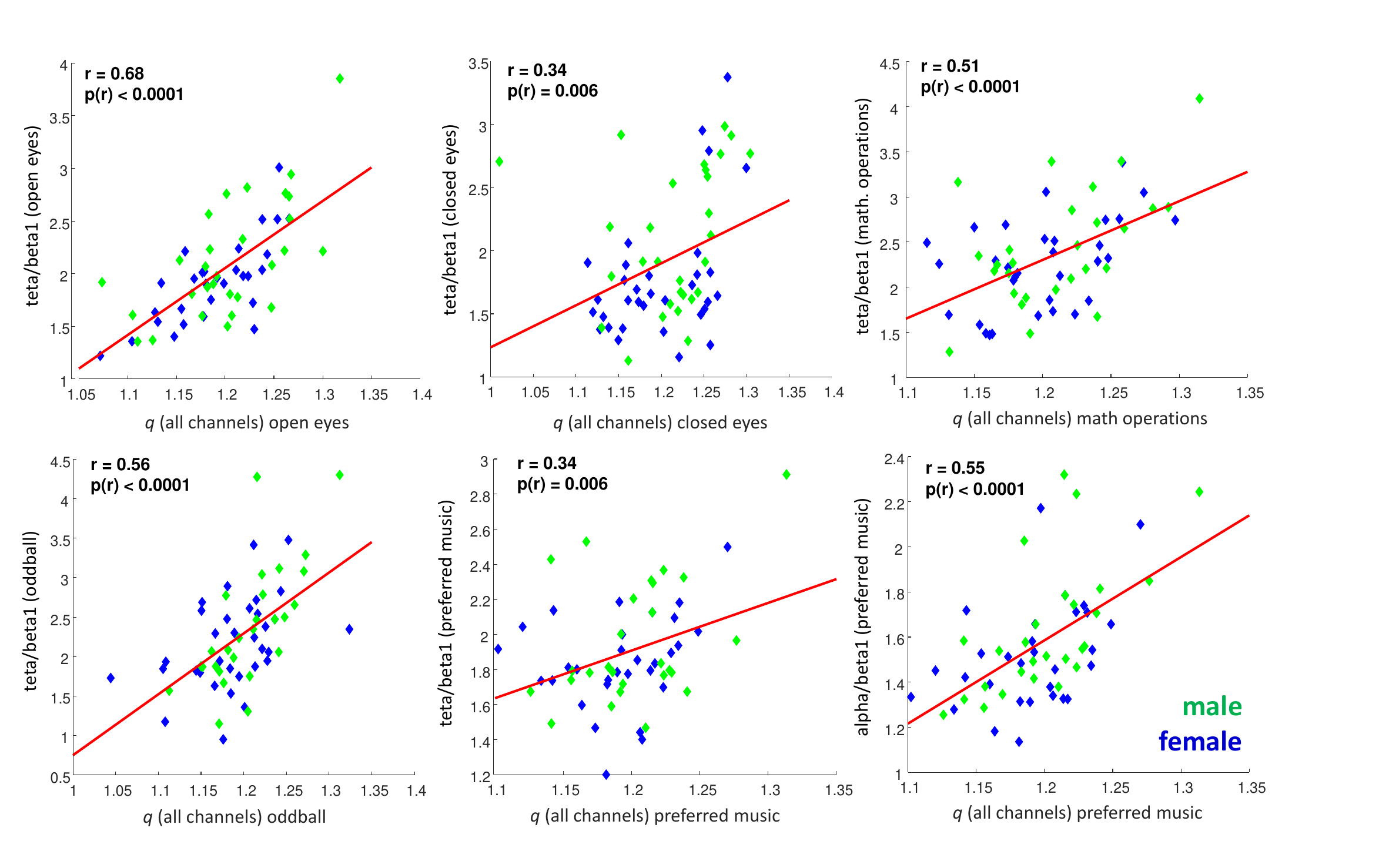}
\caption{Pearson correlations between $\theta$ (all FSs) and $\alpha$ (preferred music) relative powers and the value of the $q$ parameter (all channels). The relative powers were obtained by dividing the absolute $\theta$ (4-8 \,Hz) or $\alpha$ (8-12\,Hz) powers by the $\beta_1$ (15-25\,Hz) power. Correlation coefficients and their $p$-values are shown in each panel. The sex discrimination was illustrative.}
\label{fig7}
\end{figure}

\begin{figure}[ht]
\centering
\includegraphics[width=15cm]{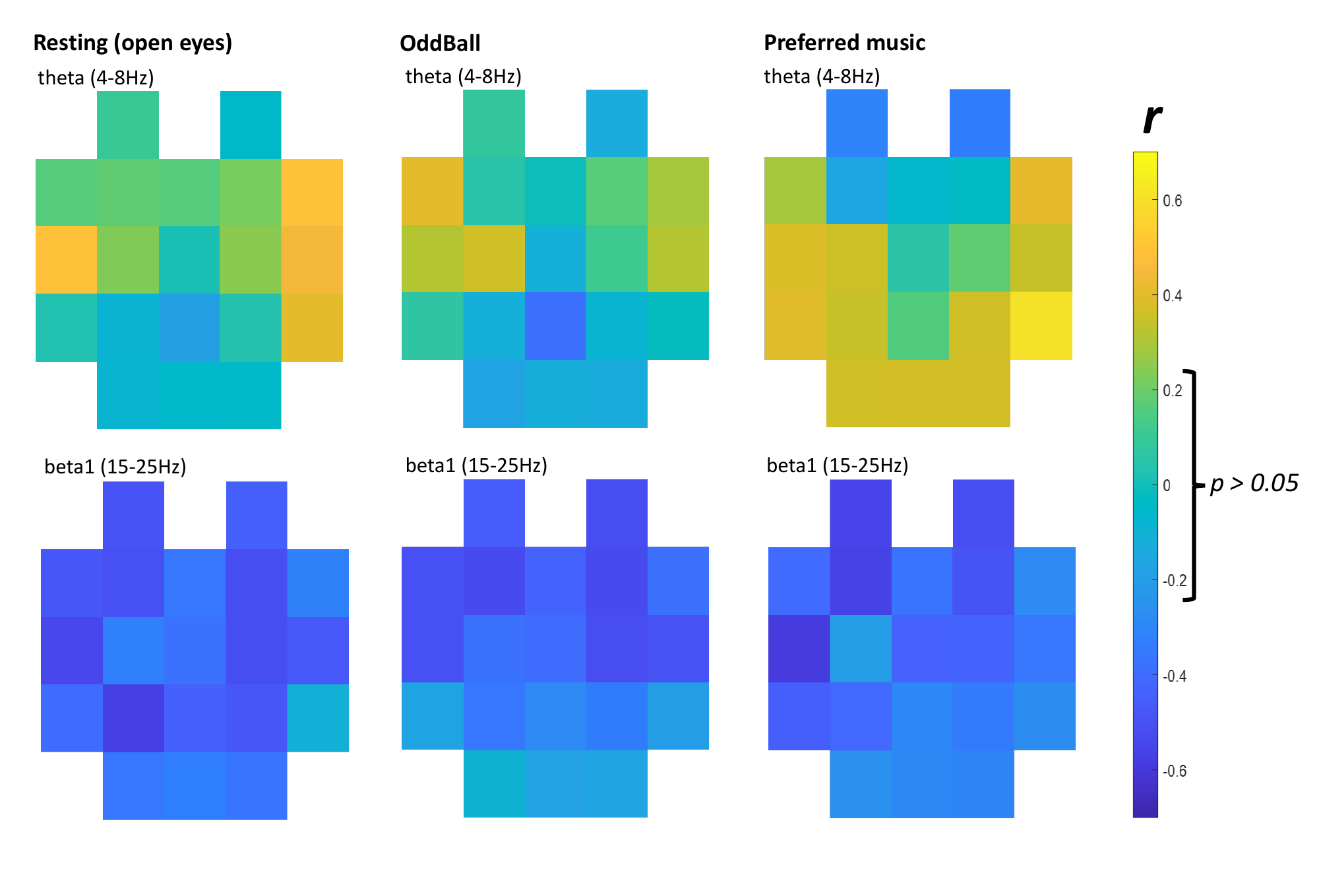}
\caption{Pearson correlation between the value of $q$ (by channel) and the power of the $\theta$ band (4-8Hz, left) and $\beta_1$ band (15-25Hz, right), by channel, in the FS of resting (eyes open) (top), Oddball (middle row), and Preferred music (bottom). The color scale encodes the values of the correlation coefficient, $r$. The range between the brackets contains r values with $p>0.05$.
}
\label{fig8}
\end{figure}

When we computed the relative powers in the $\theta$/$\beta1$ or $\alpha$/$\beta_1$  ratios, we summed up the effects of both absolute frequencies in the correlation with the $q$ parameter. In Figure \Ref{fig8}, we can observe significant positive correlations between the relative powers and FSs.

When we assessed the power by band and channel for each functional state, we generally observed positive correlations between the respective $q$ and low frequencies, and a negative correlations between $q$ and high frequencies, which reaches a Pearson's coefficient of $r =-0.65$ for $\beta_2$ in the right frontopolar channel (FP2) in OddBall FS (Supplementary Table S6, with all correlations between each band and each $q$ value, by channel and functional state). 

We selected in Figure \ref{fig8} the topographic maps with correlations between $q$ by channel and the $\theta$ and $\beta_1$ band powers for the following FS: Resting (open eyes), Oddball (where selective and sustained attention is maximally required to perform a monotonous task), and Preferred Music (when we consider to be evoking a brain multidimensional FS\cite{ Vuust2022}). All the correlations between the power of $\beta_1$ frequencies and $q$ were negative, many of them being significant. This pattern was also characteristic between $q$ and power of the $\beta_2$ and $\gamma$ frequencies in all the FSs studied (figure \ref{fig8} and Supplementary Table S6). The correlations between $\theta$ frequencies and $q$ were found to be predominantly positive and topographically more peculiar: in the conditions considered in figure \ref{fig7}, they had higher positive values in the temporal regions, which in the state of Preferred Music also extended to all posterior regions bilaterally. 

Interestingly, $\theta$ powers across single channels are higher in the Oddball paradigm than Listening preferred piece of music, being the differences significant on C3 (Oddball: 2217.60$\pm$574.09,	music: 2015.36$\pm$373.37, t-stat=2.40m, p=0.017, t-test for paired samples), C4( Oddball: 2223.38$\pm$575.86, music: 2042.98$\pm$390.74, t-stat=2.11, p=0.037), CZ (Oddball: 2704.17$\pm$632.90, music: 2368.01$\pm$423.55, t-stat=3,59 p<0,001), PZ (Oddball: 2325.95$\pm$665.21, music: 2035.98$\pm$416.24, t-stat=3,00, p=0,003) and OZ (Oddball: 1624.14$\pm$523.98	music: 1459.65$\pm$307.59, t-stat=2.20, p=0.029; see Supplementary Table S7 for all data).

\section*{Discussion}

In this study, we gathered new evidence supporting the analysis of NC through the modeling of EEG signal non-regularities using $q$-statistics \cite{NOSSO2023, NOSSO2024} while removing stereotypal thalamo-cortical activity of the PDR, predominant in the absence of visual information (eyes closed) \cite{Niedermeyer2011}. From the remaining frequency distribution, we obtain the $q$ parameter, historically considered a measure of complexity in both natural and artificial systems, ranging from physical to social ones \cite{Tsallis2023}. The value of the $q$ parameter from all channels taken together and the mean of $q$ values from single channels can evidence important nonlocal correlations.

Individual channels show different EEG signals, so we can assume that each channel somehow expresses the NC of some individual component of the brain's systemic organization, i.e. a putative complex $subsystem$ of a hierarchically lower level. Consistently, we have verified that the behavior of NC is different at global (AllCh) and local (single channels) levels, pointing to a hierarchical organization of brain networks: at the global level, the NC complexity seems to be sensitive to characteristics of the subject (as aging), and at the local level, NC can be also dynamically modulated by contextual demands of a current functional state. An important finding is the positive correlation between NC and $\theta$ power across the brain, a frequency pattern known to be related to brain integration \cite{Lisman2013, Karakas2020, Herweg2024, Lazarev2006, Tan2024}. 

Our findings in this work seem to be generically consistent with well-known data regarding the functional anatomy of the brain and its functional connectivity and can still be partially understood through the theory of functional networks, such as the Default Mode Network (DMN) \cite{Menon2023,Raichle2015}. In future studies, statistical NC dynamics can be addressed using $q$-statistics to study different mental conditions, such as mental disorders and neurodiversity, searching for possible markers of such conditions.

\subsection*{NC has different characteristics at global and local levels of brain activity} \vspace{0.5cm}

The $q$ parameter seems suitable for describing complex systems such as the human brain at different hierarchical levels of organization since this parameter systematically exhibits a behavior consistent with non-local correlations. The $q$ values from all channels (AllCh) are larger than the averaged $q$ values coming from the single channels, demonstrating that the behavior of NC is not linear amidst the parts and the whole, being larger in the latter one. 

Conversely, there was a negative correlation between age and NC, both at the global level (all channels, AllCh, Table 2) and at the local level (every single channel). Evidence shows that organic complexity, at some scale, decreases in pathological conditions and with natural aging \cite{Hatterjee2017, Goldberger2002, Goldberger2002b}, while there is no sufficient consensus on how mental disorders affect brain complexity \cite{Hernandez2023, Yang2013}. Our findings support the significance of the $q$ parameter as a measure of NC. Organic complexity might perhaps be understood as "vitality," which would decrease in the presence of disease and with aging.

While NC correlated with a general characteristic of an individual (such as age) in AllCh, the FSs had no effect on NC at the global level. This observation is consistent with the understanding that functional states, related to task performance and other transient mental circunstances, perhaps have relatively little impact on the NC of the brain as a whole. We suggest that NC at the global level may be more closely related to stable constitutional or acquired properties of human individuals (such as personality, cognitive profile, training, chronic stress exposure, mental disorders, etc.) than transient task-related FSs. This hypothesis aligns with various evidences in the literature \cite{  Hernandez2023,Yang2013,Aguerre2023, Gu2024,  Gu2022,Hadoush2019, Lau2022} and with unpublished results obtained by our team. 

However, NC in the EEG from single channels was significantly affected by FSs (Figures \ref{fig5} and \ref{fig6}), indicating the occurrence of modulation in the subsystems operating during the tasks/procedures in the seven FSs of this study. This modulation seems consistent with current models of brain physiology, considering different neural components (or subsystems) accessed by each EEG channel. However, using setups with 20 electrodes \cite{Seeck2017},the estimation of the brain sources of the signals captured by the EEG channels cannot be determined. Therefore, any inference made here about the signal source is roughly estimated.

A marked effect of closed eyes was observed (even more pronounced during photic driving), reducing neural complexity in the occipital, parietal, and posterior temporal subsystems bilaterally (Figures \ref{fig4} and \ref{fig5}, corresponding panels), which aligns with the reduction of visual information. The PDR in the $\alpha$ frequency band, a steady and highly regular oscillation, did not interfere with this NC measure once it was removed. Informativeness exerts effect on NC, what seems to  be evidenced here.

Despite median fronto-central NC was not statistically different across FSs (Figure \ref{fig4}), the NC from FZ and CZ was more expressive than NC from almost all other sites with eyes open, and NC was reduced compared with other channels in the FSs with eyes closed (Figure \ref{fig5}). Conversely, we observed an increase in NC in the frontal polar channels (FP1 and FP2) during closed eyes when comparing NC across states (Figure \ref{fig4}). In all FSs, frontotemporal sites (F7 and F8) showed significantly higher NC compared with almost all other EEG sites (Figures \ref{fig4}, \ref{fig5} and \ref{fig6}). These six channels can record the activity (and NC) mainly of the frontal subsystems in general, which are related to the internal work in economic judgment, semantic and emotional/motivational evaluation of complex contexts (such as social ones), reasoning, behavioral planning and decision making. \cite{Hogeveen2022,Kringelbach2004,  Mansouri2017, Pana2023, Wojtasik2020}. These NC patterns may be a reflection of the modulation of the activity from frontal regions, which specifically and differentially participate in different types of functional networks, such as the DMN \cite{Menon2023,Raichle2015}. 

While listening to preferred piece of music, a complex experience may be elicited as a complex sensory, mnemic and emotional high-order representation. This type of state would be compatible with the relative increase in NC in posterior regions, which are critical for subjective experience of memory retrieval \cite{Foudil2024, Simons2022}, sensory cognition \cite{Cohen2009,Mesulan1998}, as well auditory processing \cite{Leech2011}, which could be very complex regarding this musical experience. The posterior increase in NC appears to have occurred asymmetrically, as the right temporal and parietal subsystems were more prominent when compared to their contra-lateral counterparts. Cognitive processing of music seems to be related to neural networks in the right hemisphere \cite{Gunturkn2019, Zatorre2022}, and there is also a classic model that attributes the primary role in processing emotions to the right hemisphere \cite{Gunturkn2019}.

\subsection*{Theta waves are correlated with neural complexity}\vspace{0.5cm}

We observed that, in general, NC measured by $q$-statistics is positively correlated with slow-frequency bands, especially the $\theta$ band (4–8 Hz), and negatively correlated with fast-frequency bands $\beta_1$ (15–25 Hz), $\beta_2$ (25–35 Hz), and $\gamma$ (35–50 Hz). The relationship between spectral composition and NC was consistent when analyzing both AllCh and single channels.

Either episodic or semantic declarative memories are high-order representations formed and evoked through emergent processes of association between perceptions, cognitions, emotions/motivations, and other pre-established memories, widely recruiting brain subsystems in synchrony through functional connections in the $\theta$ frequency, distributed across the brain from the medial temporal lobes \cite{Lisman2013, McNaughton2022, Herweg2024}. Through phase locking between local $\gamma$ frequencies, which probably provides processing of specific and local information in each subsystem, and $\theta$ frequency that apparently reflects propagation of signals over long distances in the brain from the medial temporal lobe, functional connectivity and the integration of a myriad of pieces of information occur to form a high-dimensional representation \cite{Lisman2013}. The integrative and associative role of the $\theta$ band has long been well-established, though few studies directly correlate it with NC. However, they have linked complexity in the $\theta$ band with its integrative role via long-range functional connections \cite{Gonzalez2022,Stokic2012}. One such study associated an increase in $\theta$ band complexity (3.5–7.5 Hz) with monochord sounds \cite{Bhattacharya2016}.

Given that the $\theta$ band is intrinsically related to the physiology of the temporal lobe structures \cite{Buzsaki2002, Pignatelli2012}, we found significant correlations between the $\theta$ band and NC in temporal channels, bilaterally. These correlations were less evident in the Oddball functional state, where informativeness is presumably lower, limited to cognitive resource allocation to sounds discrimination in a monotonous task, and of greater magnitude in resting (with eyes open) and preferred music FSs, when the individual would freely express complex associations in their internal cognitive life. In the preferred music FS, the higher correlation between NC and the theta band spreads from temporal channels to frontotemporal (F7, F8), lateral central (C3, C4), lateral parietal (P3, P4), occipital, and occipitotemporal (T5, T6) channels, bilaterally. Preferred music may evoke high-order representations with cognitive and emotional components ($qualias$), consistent with the observed increase in the correlation between complexity and the $\theta$ band. However, we found slighty higher $\theta$ power in Oddball than in Preferred Music in the EEG of almost all single channels. These findings suggest a dissociation between the power and the NC of the $\theta$ activity, because complexity is related to both functional connectivity and the amount of information in the system. An epileptic seizure is a state of absolute correlation between the components of the brain since there is absolute synchronization of the brain's neural networks, manifesting in the EEG as high-amplitude standing waves. In this example, there is no NC because there is no information in the neural networks nor any expression of the singularity of their components. Thus, the power of the $\theta$ band is not sufficient to determine complexity, justifying lower NC during the Oddball paradigm, where brain resources must be synchronized and recruited for a task that is (as many participants said) monotonous and simple.

The powers of frequencies above 15Hz were negatively correlated with NC across all brain channels, regardless of the state. We have shown data related to $\beta_1$ frequencies in Figures \ref{fig7} and \ref{fig8} $\beta_2$ $\gamma$. Gamma bands manifested the same general pattern of correlations with NC. This finding can be interpreted as representing higher frequencies involved in local information processing \cite{Niedermeyer2011, Lazarev2006, Lisman2013}, while NC is related to long-range interactions. It is believed that the amplitude desynchronization of $\beta$ activity during mental activity is a subtle indicator of local processes of cortical activation during task performance \cite{Fernandez1995, Pfurtscheller1999, Lazarev2006}. For fairly long epochs of analysis (from seconds to many minutes, as in the present study), the absolutely predominant part of the time the normal EEG is characterized by desynchronized low-amplitude $\beta$ waves with multiple and varying frequencies typically associated with active wakefulness and busy, or anxious thinking and active concentration \cite{Baumeister2008, Grafton2012}. However, transient brief high-power $\beta$ bursts seem to reflect a large-scale brain integration that has an inhibitory nature, filtering out irrelevant information for the task at hand \cite{Lundqvist2024}. The $\beta$ synchronization, manifesting higher power amplitude, e.g. during drowsiness and sleep \cite{Vakulin2016} or in use of barbiturate or benzodiazepines \cite{Feshchenko1997, Nishida2016}, is related to local inhibitory processes \cite{Pfurtscheller1992}. Perhaps as a corollary, the $\beta$ band would sustain and stabilize the current sensorimotor or cognitive states \cite{Lundqvist2024, Engel2010}, and possibly ranging to inflexibility in pathological states \cite{Engel2010}. Either short-range processing by $\beta$ and $\gamma$ frequencies, as low informative long-range functional connectivity expressed by $\beta$ band would be compatible with lower NC, explaining our findings about higher frequencies.

\section*{Final considerations}

The approach to neural complexity in the typical, atypical, and altered brain through $q$-statistics and the associated psychosocial and cultural universe promises a productive line of research. The present paper is a exploratory study about physiological aspects of Neural Complexity in relation to neural patterns across seven different states and tasks, which are mentioned here as “Functional State” (FS). From here, many new questions arose, which need specific experimental designs to clarify them, as high density EEG for source estimation or techniques of functional imaging, for example. 

We initially sought to reaffirm the strong relationship between the parameter $q$ (an index which qualifies the nonadditive entropy) and the NC demonstrated in our previous works \cite{NOSSO2023, NOSSO2024}, highlighting some evidences of nonlocal correlations behavior of $q$ and confirming the inverse relationship between age and NC, consistently with what is available in the literature. 

Subsequently, we studied the behavior of brain complex subsystems through parameter $q$, which was obtained from the EEG of the single channels. In contrast to the higher level of organization (AllCh), the FS play an important role in modulating the NC of brain subsystems, which was coherent with loss of information in visual areas with eyes closed and also could be related to evoking different brain networks. We showed higher correlations of NC with the $\theta$ band in the FSs of higher informativeness, linking $q$ to the integrality and also to the informativeness of a system that establishes a network of long-range and intricate correlations probably manifesting high-order representations.
 
For future publications, we are in parallel studying the relationships between neural complexity inferred by $q$-statistics and cognitive characteristics of individuals (measured by WAIS) and clinical characteristics estimated by the scales, as well as the potential of $q$-statistics for precision neuro/psychodiagnostics through $(c,q)$ maps \cite{NOSSO2024}. These topics are broad, promising, and deserve to be suitably addressed.

\section*{Data availability}
The database analyzed in this work is in Supplementary Dataset S8. The Raw EEG data will be made available on request.

\section*{Acknowledgments 
}

We thank Lucas Teles da Silva, Maria do Socorro Santana and Aldenys Perez for useful remarks. This academic research was supported by the Programa de Incentivo a Pesquisa - IFF/FIOCRUZ-FIOTEC. One of us (CT) acknowledges partial financial support of the Brazilian agencies CNPq and Faperj.

\section*{Author contributions statement}

C.T. developed the mathematical model, D.M.A., C.T., and V.V.L. conceived the experiment(s),  D.F.Q, R.P.C., C.K.L. conducted the experiment(s), D.M.A., R.P.C, D.F.Q., and H.S.L. analyzed the results.  D.M.A. wrote the manuscript. All authors reviewed the manuscript. 

\section*{Competing Interests}
The authors declare no competing interests.

\section*{Additional information}
Correspondence and requests for materials should be addressed to D.M.A.



\end{document}